\documentclass[twocolumn,prl,superscriptaddress,floatfix,showpacs]{revtex4-1}
\usepackage[utf8]{inputenc}
\usepackage{amsmath}
\usepackage{amssymb}
\usepackage{graphicx}

\makeatletter
\usepackage{graphics}
\usepackage{epsfig}
\usepackage{color}

\usepackage[normalem]{ulem}
\newcommand{\be}{\begin{equation}} \newcommand{\ee}{\end{equation}}
\newcommand{\bea}{\begin{eqnarray}} \newcommand{\eea}{\end{eqnarray}}

\makeatother

\begin{document}

\title{Morphological transitions in supercritical generalized percolation and moving interfaces in media 
with frozen randomness}

\author{Peter Grassberger} \affiliation{JSC, FZ J\"ulich, D-52425 J\"ulich, Germany}

\date{\today}

\begin{abstract}

We consider the growth of clusters in disordered media at zero temperature, as exemplified by supercritical 
generalized percolation and by the $T=0$ random field Ising model. We show that the morphology 
of such clusters and of their surfaces can be of different types: They can be standard compact clusters 
with rough or smooth surfaces, but there exists also a completely different ``spongy" phase. Clusters 
in the spongy phase are `compact' as far as the size-mass relation $M \sim R^D$ is concerned (with 
$D$ the space dimension), but have an outer surface (or `hull')
whose fractal dimension is also $D$ and which is indeed dense in the interior of the entire cluster.
This behavior is found in all dimensions $D \geq 3$. Slightly supercritical clusters can be of either 
type in $D=3$, while they are always spongy in $D\geq 4$. Possible consequences for the applicability 
of KPZ (Kardar-Parisi-Zhang) scaling to interfaces in media with frozen pinning centers are studied 
in detail. In particular, we find -- in contrast to KPZ -- a ``weak-coupling'' phase in 2+1 dimensions.

\end{abstract}
\maketitle

\section{Introduction}

Rough surfaces and interfaces have a huge number of applications in nature and in technology. Accordingly,
there is a very large number of papers devoted to them in the physics literature, as well as several 
monographs \cite{KPZ,Meakin,Barabasi,Family,Vicsek}. For statistical physicists, one of the major reasons
for studying them is the fact that they typically show anomalous (fractal) scaling, with deep connections 
to phenomena like phase transitions, critical points, and the renormalization group.

One very important distinction in the theory and phenomenology of rough surfaces is between growing and
pinned ones. Pinned rough surfaces can only occur in media with frozen randomness at zero temperature, 
while moving rough surfaces can be found both in random media and in ordered media with temporal 
(thermal or a-thermal) disorder. 

The prototypical model for driven interfaces without frozen disorder is the Kardar-Parisi-Zhang (KPZ) 
model \cite{KPZ}, a discrete version of which is the Eden model for cancer growth
\cite{Eden}. When starting from a flat interface at time $t=0$, it never forgets the initial growth
direction in any dimension, so that the interface never becomes isotropic either globally or locally
(this is, e.g., different for diffusion limited aggregation, where interfaces become locally isotropic
for $t\to\infty$ \cite{Hegger,Lam}). Thus growing interfaces in non-random media are {\it self-affine}.

This is not always the case for pinned interfaces. In two dimensions, it is shown in 
\cite{Drossel,Grass2018} that critically pinned interfaces in isotropic random media are always in
the universality class of critical percolation and thus fractal and isotropic on large scales. More 
precisely, the interfaces are in the class of percolation {\it hulls}, i.e. of externally accessible 
surfaces of critical percolation clusters. In $D\geq 3$, there are two different universality 
classes for critically pinned interfaces, depending on control parameters (plus a tricritical region 
for intermediate control parameters) \cite{Ji,Koiller,Janssen,Bizhani}. One of them is again the 
percolation universality class, while the other is believed to be self-affine 
\cite{Meakin,Barabasi,Family,Koiller} -- although there are numerical indications that this might 
not be strictly true, in particular in the weak disorder limit \cite{Grass_unpub}. The tricritical 
point separating these two regimes was studied by renormalization group method in \cite{Janssen},
and numerically in \cite{Bizhani}. While the agreement between the two is far from perfect in $D=3$, 
at least the locations of the tricritical points are known with high precision for all 
$D < 6$ \cite{Bizhani}.

In the present paper we shall deal with the case of moving interfaces in (isotropic) media with 
frozen disorder. Notice that any frozen disorder leads, for sufficiently weak pushing, to pinning
sites where the interface stops moving, while it continues to move globally. There is a wide spread 
believe that they should also be described by KPZ 
scaling \cite{Meakin,Barabasi,Family,Vicsek}, because the `frozen-ness' of the noise should not be 
very relevant as long as the interface moves. It might occur occasionally that a particularly 
strong obstacle prevents the interface from progressing locally, in which case the interface will
stop to grow locally and close again behind the obstacle. This leaves then a bubble, but these 
bubbles should not modify the basic scaling laws. Notice that most early experiments which looked
for KPZ scaling in real phenomena \cite{Rubio,Horvath,He} indeed involved frozen 
randomness.

We should stress that, as soon as (local) pinning is possible, the mapping of the KPZ problem onto
the directed polymer model \cite{Halpin} breaks down, as the polymer would get adsorbed at the 
pinning center. Indeed, there is no theoretical argument why interfaces with local pinning should be
in the KPZ universality class, although is is generally assumed \cite{Meakin,Barabasi,Family,Vicsek}.

An essential aspect of this standard scenario is that overhangs of the interface can be neglected
in the scaling limit, i.e. the interface can be described by a single-valued height function $h(\bf x)$.
If there is no frozen randomness, i.e. in the KPZ case proper, this seems to be correct (although I am 
not aware of a rigorous proof). Detailed models of interfaces without overhangs (e.g. in so-called
solid-on-solid (SOS) models) in the presence of frozen randomness have been studied in the literature
\cite{Bruinsma,Huse-Henley}, but the assumption that overhangs can indeed be neglected has, to my knowledge,
never been challenged seriously.

As we shall see, it is indeed wrong. In some control parameter regions this scenario is completely
and qualitatively overthrown, due to the existence of what we call a `sponge phase', where the interface
is even more rough than a fractal. In the sponge phase, the interface is indeed dense in the entire bulk
underneath the surface, with deep fjords reaching from the top down to the bottom nearly everywhere. 
The standard scenario is still true qualitatively in other regions, but whether the KPZ
scaling laws hold there is not completely clear. In order to study this, we have to define a suitable
``effective interface'', where the deep fjords in the sponge phase are cut off. Using this, we 
find clear evidence that there exists a ``weak-coupling'' phase in $D=3$ (i.e., 2+1) dimensions, while 
such a phase exists for KPZ only for $D>3$ \cite{Nattermann}. In the strong coupling phase, we find 
critical exponents roughly in agreement with KPZ.

In the next section we shall define the model which we used for our numerics, but we stress that the 
phenomena discussed should hold much more generally. In Sec.~3 we present theoretical arguments for 
what we call {\it sponge phases} and numerical results supporting them. In Sec.~4 we discuss whether 
KPZ scaling holds in non-spongy phases -- and maybe even in sponge phases, provided one modifies
suitably the definition of interfaces. In Sec.~5 we summarize and draw our conclusions.

\section{Generalized percolation}

We consider (hyper-)cubic lattices in dimensions $D = 3$ to 7 . The class of models we study can
be considered as a generalization of the susceptible-infected-removed (SIR) epidemic model, where 
we keep track of how often a (not yet infected) site had been ``attacked" (and thus maybe weakened) 
by an infected neighbor \cite{Janssen}. 
Thus each lattice site can be either `removed' (i.e., it had been infected, but it no longer is), 
infected, or in one of ${\cal N}+1$ susceptible states, where ${\cal N}$ is the coordination number. 
Time is discrete, and updating is done in parallel:
At each time steps, every infected site attacks each of its neighbors, after which it becomes removes.
If a neighbor had already been attacked $k-1$ times (either previously or during the present time step)
but is still susceptible, it becomes infected with probability $p_k$ \cite{Janssen}. Thus, after $k$ 
attacks it will have been infected with probability $q_k$ with
\be
    q_k = q_{k-1}+(1-q_{k-1}) p_k .
\ee
We shall call this model {\it generalized percolation}. For efficiency of the code, all infected 
(or `growth' or `active') sites are written in a list, and during each 
time step this list is gone through, a new list of growth sites is built, and at the end the old list is 
replaced by the new one. Notice that resulting configurations might depend on the order in which
these lists are gone through, for a fixed sequence of random numbers. Therefore, after each time step 
the list of growth sites is randomly permuted.

Special cases of this model are site and bond percolation. In the former, a site cannot be infected at 
all, if the first attack did not already succeed. Thus $p_1=p$ and $p_k = 0$ for $k\geq 2$. For bond 
percolation, in contrast, $p_k$ is independent of $k$: $p_k = p$ for all $k$.

Consider now the random field Ising model (RFIM), with an initial state where all spins are down
except for `seed sites', whose spins are up. Dynamics is single flips with parallel update, and we 
assume that that the spin at site $i$ can only flip if (i) this reduces the energy (i.e., we are at $T=0$), 
and (ii) at least one of it neighbors have flipped during the previous time step. It is easily seen
\cite{Drossel,Grass2018} that this can be mapped exactly onto the above generalized percolation model,
for any distribution of local fields.

Alternatively, one could consider a porous medium that is wetted by a fluid with non-zero surface
tension. Although one now cannot prove a general mapping, it is clear that generalized percolation
will be a good model for a wide range of local geometries and surface tensions.

\begin{figure}
\begin{center}
	   \vglue -.4cm
   \includegraphics[width=0.49\textwidth]{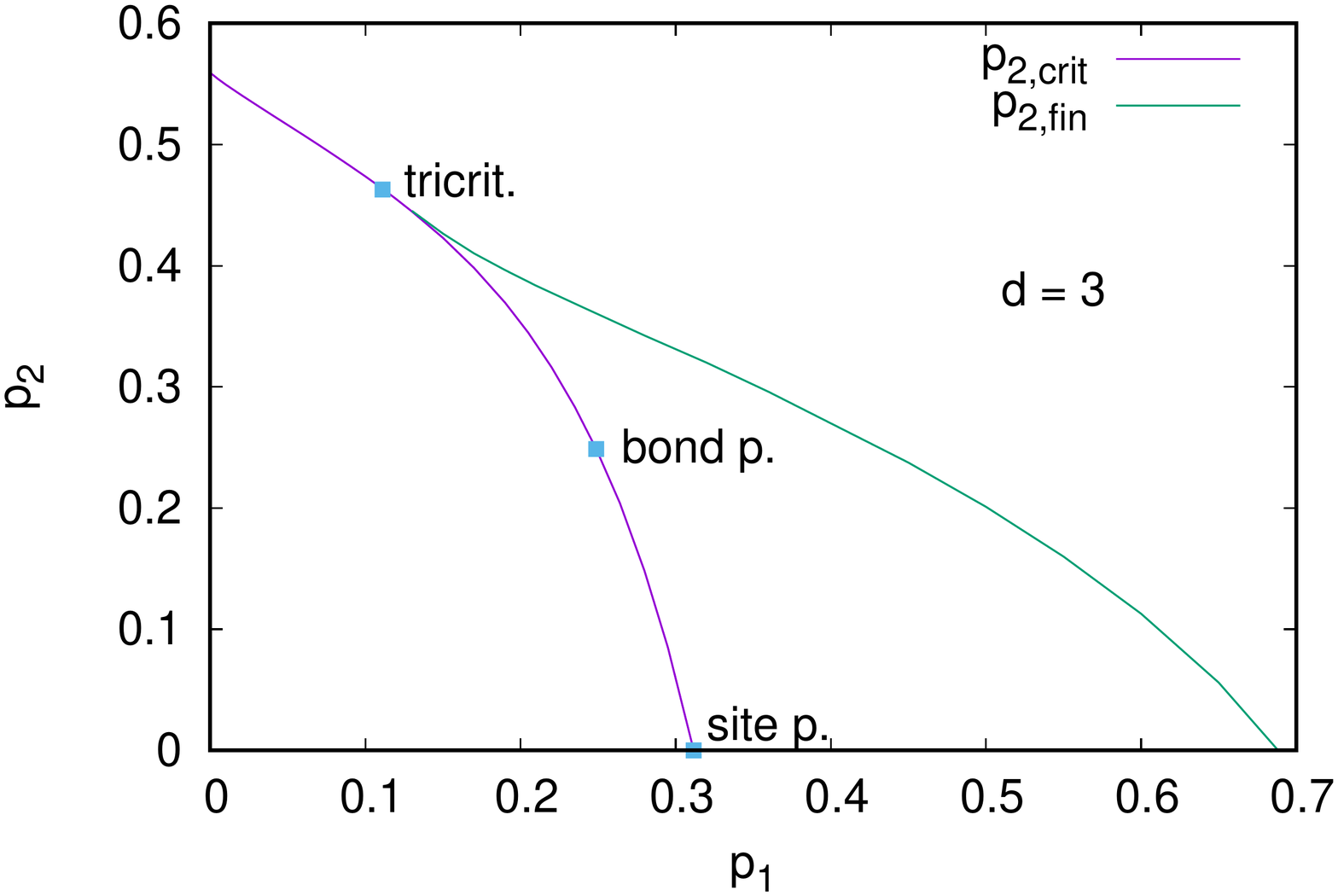}
   \vglue -1.0cm
   \includegraphics[width=0.49\textwidth]{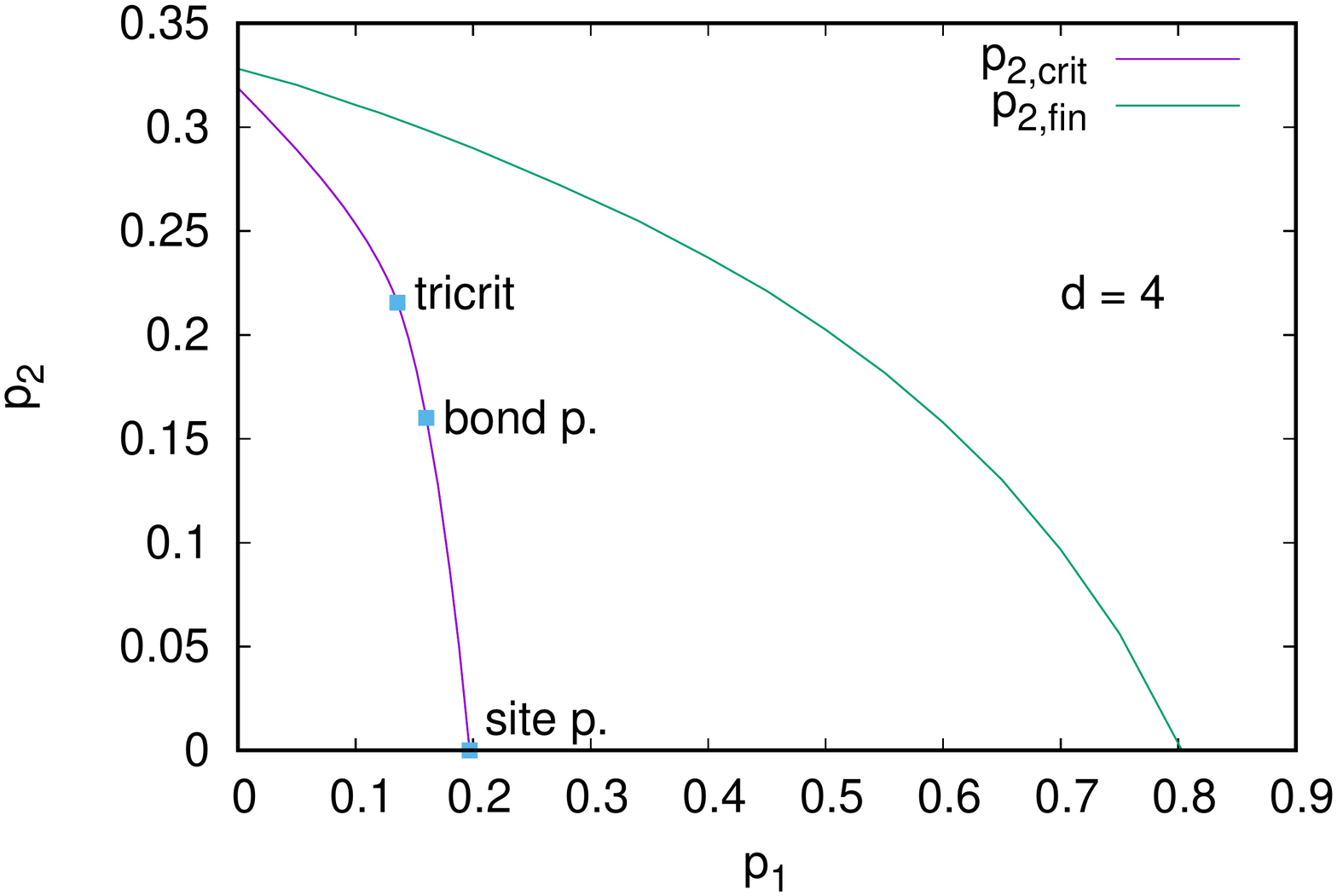}
   \vglue -1.0cm
   \includegraphics[width=0.49\textwidth]{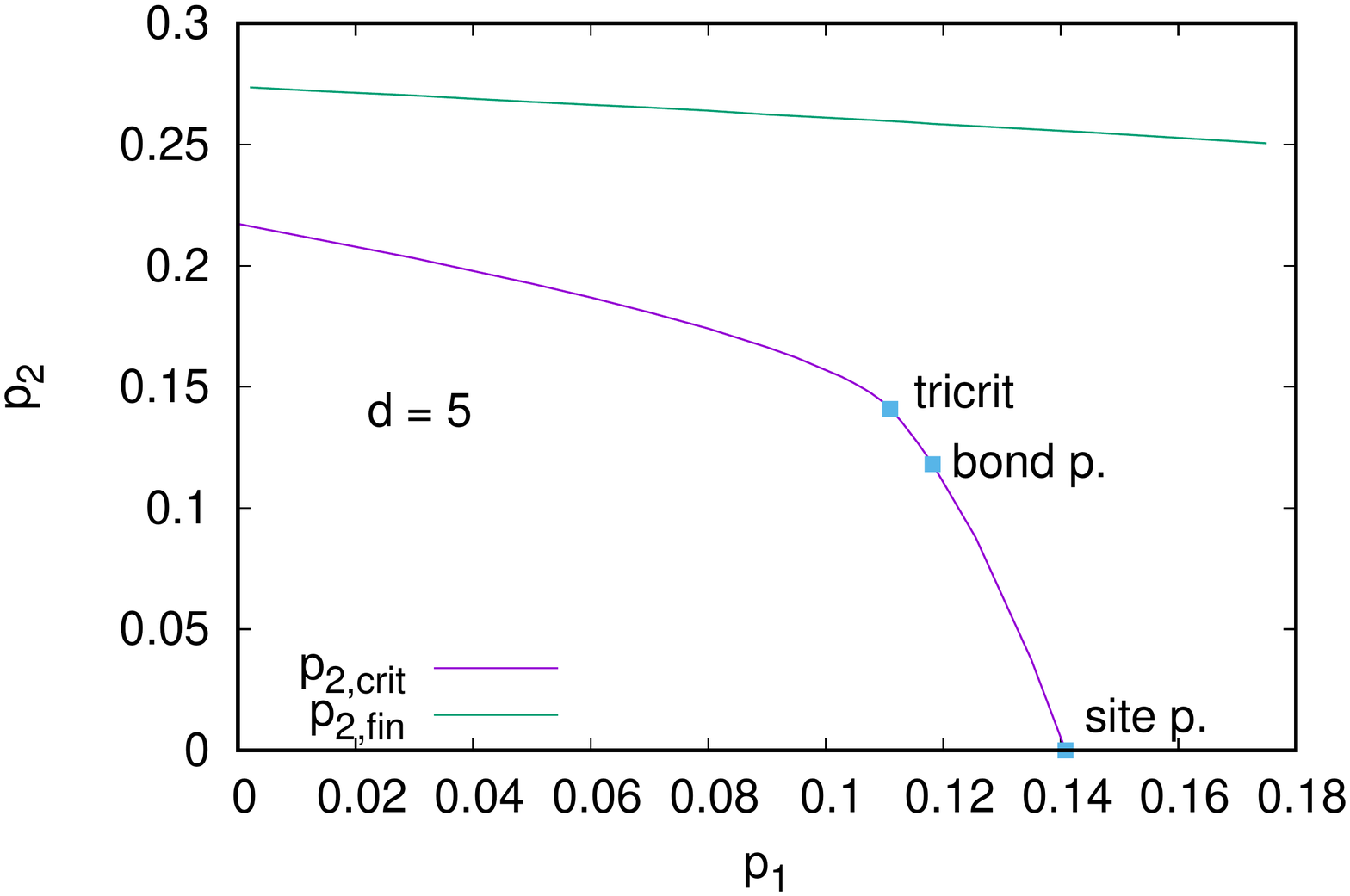}
   \vglue -1.3cm
\end{center}
	\caption{(color online) The lower curves in each panel represent the critical curves (data 
	from \cite{Bizhani}). Below them, clusters starting from point seeds are finite and interfaces 
	move only a finite distance, before they stop. Above them, interfaces would grow forever on 
	infinite lattices, and clusters starting from point seeds would have a finite probability to 
	grow forever. On these critical curves are indicated the tricritical points \cite{Bizhani} 
	and the critical bond and site percolation points. Critical
        interfaces are fractal to the right of the tricritical points, while they are rough on their 
	left sides. The upper curves separate, for $D=4$ and 5, the sponge phase (between both curves) 
	from the non-spongy phase. For $D=3$, the same is true to the right of the tricritical point. 
	On the left of the tricritical point, there is no sponge phase for $D=3$.}
\label{fig:phases}
\end{figure}

There are strong reasons \cite{Ji,Koiller,Janssen,Bizhani,Grass2018} to believe that there are only 
three universality classes for critical generalized percolation. One is ordinary percolation. The 
second is found in cases where first attacks are very unlikely to succeed, but lead to much weakened
sites. Thus $p_k$ increases strongly with $k$. In that case, clusters tend to be less fractal and 
their surfaces tend to be more smooth. Indeed, `bays' and `fjords' will become infected (because 
sites there have many infected neighbors), while tips and spikes will be formed less likely. This 
is precisely the case for wetting of porous media by a fluid with high surface tension, and for 
the RFIM with very small randomness of the local fields. Indeed, it was found in \cite{Ji,Koiller}
that there is a sharp morphological transition in critically pinned interfaces in porous media and
in random magnets, where the interface changes from percolation-like at small surface tensions
and strong disorder to rough but non-fractal at large surface tension and weak disorder.
Between these two regimes is a tricritical point \cite{Janssen,Bizhani}. Essentially, we shall
in the present paper deal with analogous (but very different in details) transitions for 
non-pinned interfaces.

In view of our claim of universality, we shall study only the simplest non-trivial model 
for generalized percolation, which was first studied in \cite{Bizhani} and was called the `minimal 
model" (MM) in \cite{Grass2018}. There, we assume $p_1 \neq p_2$ but $p_k = p_2$ for all $k >2$. 
Thus, we distinguish between virgin sites and sites that had already been attacked by an infected 
neighbor, but we do not keep track of the number of previous attacks.

In all dimensions, we follow the upward evolution of an interface that was initially equal to 
the (hyper-)plane $z=0$, on lattices that were big enough so that the upper boundary of the lattice
was never reached during the simulation. Lateral boundary conditions were helical. The critical
lines in the $p_1$ versus $p_2$ control parameter space for $D=3,4,$ and 5 are shown in Fig.~1.

\section{Multiple percolating clusters and sponge phases}

Let us first consider site percolation in $D\geq 3$. For $D=3$ the percolation threshold on the SC 
lattice is at $p_c = 0.3116\ldots$, and for $D\geq 4$ it is even smaller \cite{Stauffer}. Thus $p_c$ 
is in all these dimensions less than 1/2. Take now a hypercubic lattice and color its sites randomly 
black and white, each with probability $p=1/2$. Then both the black and the white sites will 
percolate, i.e. we have at least two coexisting percolating clusters. Indeed, since the density of 
black sites is supercritical, there will be exactly one infinite black cluster, all other black 
clusters will be small. Similarly, there will be precisely one infinite white cluster. These two 
infinite clusters will then penetrate each other. If site $i$ is black and is on the infinite 
cluster, there is a finite probability that one of its neighbors is white and on the white infinite 
cluster. Finally, the same will be true not only for $p=1/2$, but also if sites are 
black with any probability $p \in (p_c,1-p_c)$ and white with probability $1-p$.

Consider now a `cluster' of black sites grown -- for a finite time $t$, and for $p \in (p_c,1-p_c)$ -- 
by starting from an infected hyperplane $z=0$ that can infect its upper neighbors \cite{footnote1}. 
Otherwise said, 
consider the set of all black sites (in a configuration with a fraction $p$ of black sites) which 
are connected to the bottom of the lattice by a path of length $\leq t$. Roughly, they will occupy 
densely a layer of thickness $\propto t$. Similarly, when starting from a point seed, the cluster 
will occupy densely a (hyper-) sphere of radius $\propto t$. By `densely' we mean that this layer 
or sphere has no {\it big} voids (near every point in the layer/sphere there is a point belonging 
to the cluster), but it contains many holes, and these holes are not only connected, but also dense. 
Indeed, the white infinite cluster will also be dense everywhere, so that the black cluster has 
fjords that penetrate it all the way down to the bottom $z=0$. 

We call the phase with two clusters penetrating each other and being dense everywhere a {\it sponge} phase.
While the above arguments clearly show  (even if not in a strict mathematical sense) that a sponge phase 
exists in site percolation for $D>3$, the situation is much more subtle for bond percolation. But 
the existence of multiple clusters was proven rigorously at least for $D\geq 8$ by 
Bock {\it et al.} \cite{Damron0}, and their denseness was proven in \cite{Damron1}.

In order to understand the situation for bond percolation with all $D\geq 3$ and for generalized percolation, 
we used Monte Carlo simulations. We studied mainly the planar geometry. Thus in a first step, we started with a
seed consisting of the hyperplane $z=0$ and let the cluster grow until the layer $z=L_z-1$ was reached for the
first time. At that moment, the growth was stopped, and in a second step a cluster was grown on the not yet
infected sites by starting from a (hyper-)planar seed at $z=L_z$ and moving down. Alternatively, in order to 
speed up the simulation, we did not grow the entire second cluster but just followed the external hull of the
first cluster (by which we mean the set of sites which are not on the first cluster but are neighboring it, 
and which are accessible by paths starting at $z=L_z$ and avoiding the first cluster).

\begin{figure}
\begin{center}
   \includegraphics[width=0.49\textwidth]{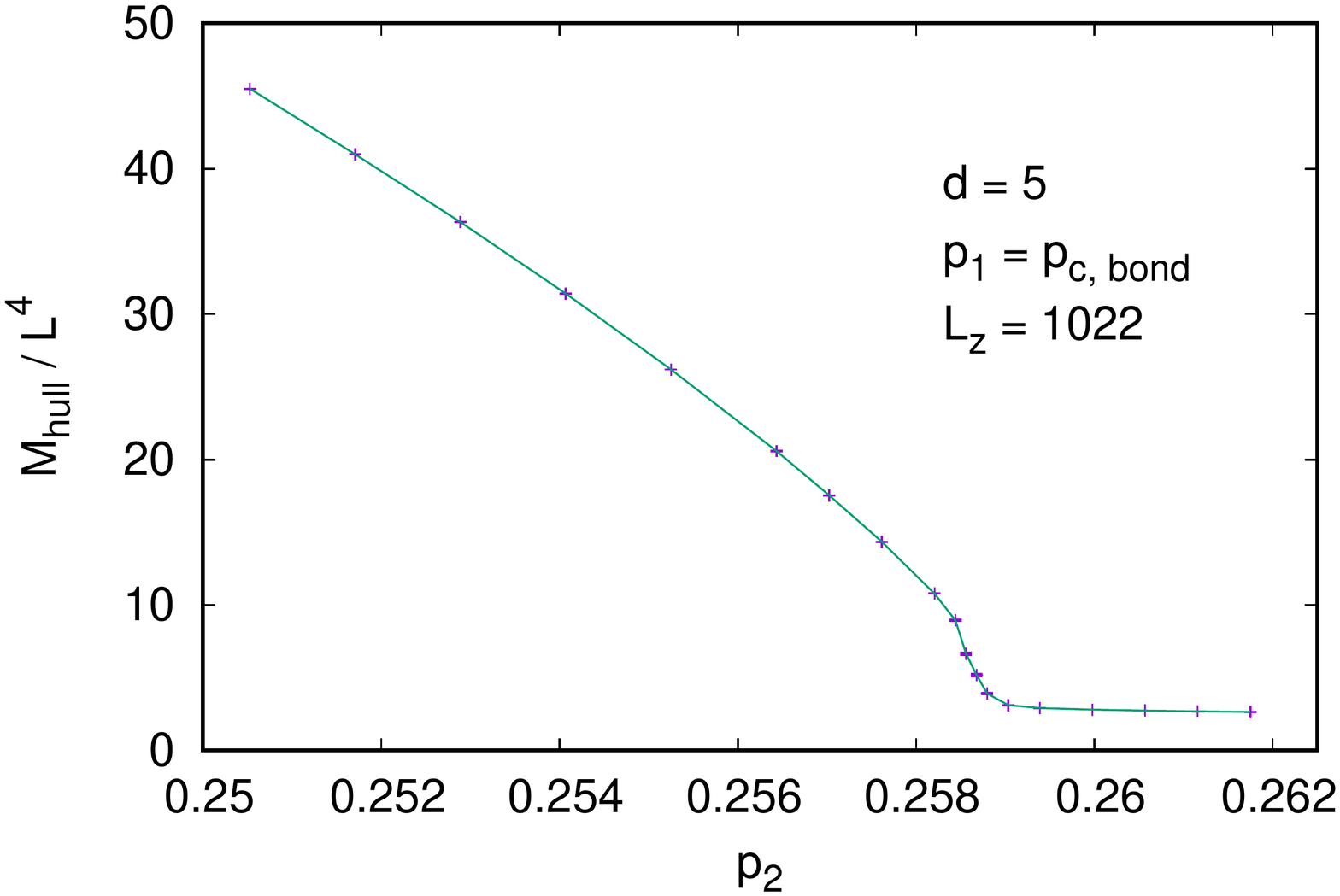}
   \vglue -1.0cm
   \includegraphics[width=0.49\textwidth]{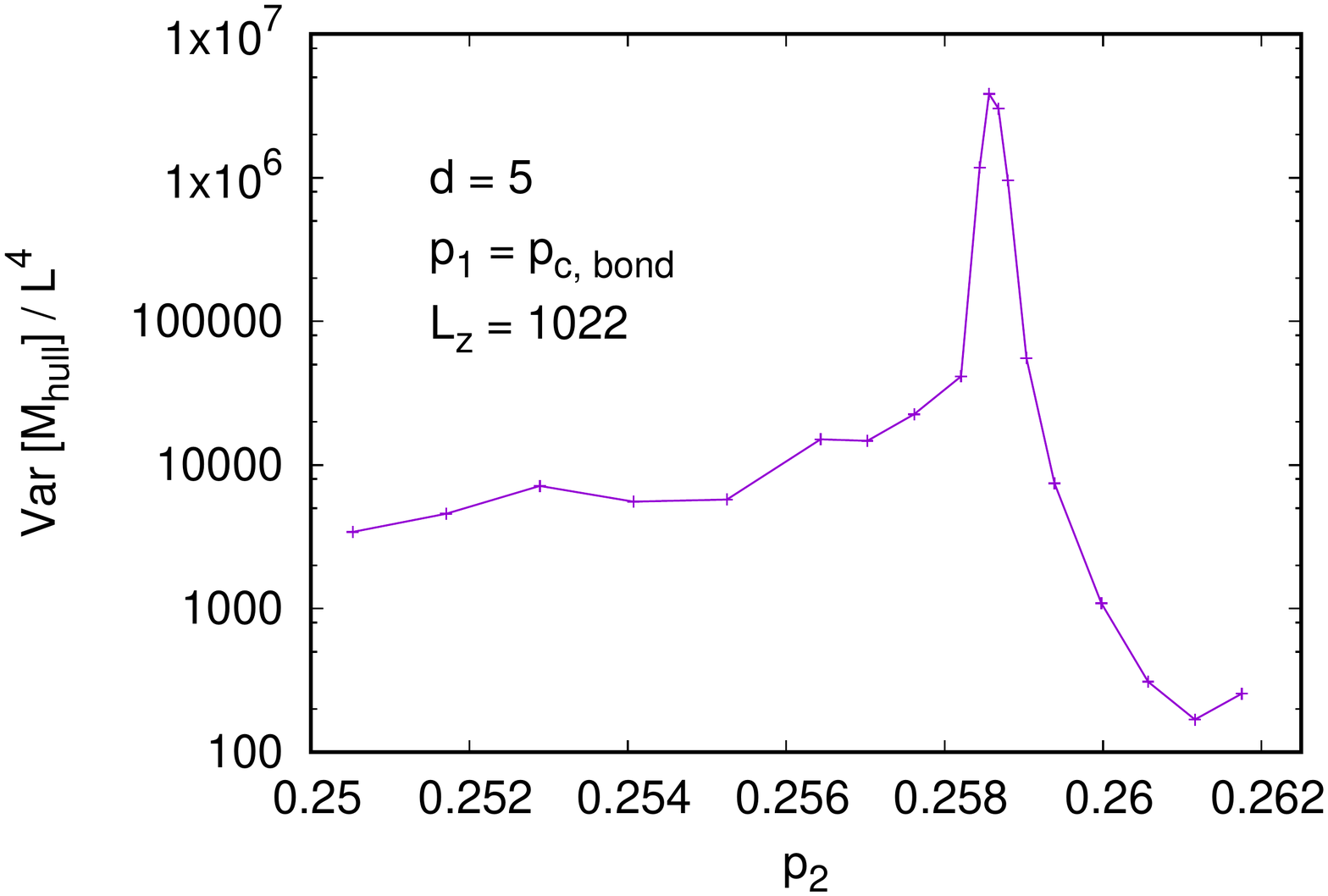}
   \vglue -1.3cm
\end{center}
	\caption{(color online) Panel (a): Mass density (per unit base surface) of the external hull of the cluster 
	growing upwards from $z=0$ in dimension $D=5$ and at $p_1 = p_{c,\rm bond}$. This density would be 
	infinite in the sponge phase, if the cluster thickness $L_z$ were infinite. Panel (b) shows the 
	variance of the mass density.}
\label{fig:hull_5D}
\end{figure}

As observables we measured the penetration depth of the second cluster, its density profile, the mass of the 
hull, and its variance. In general, the lateral size $L$ of the lattice was somewhat smaller than $L_z$.

As typical results we show in Fig.~2a the average mass density of the hull and in Fig.~2b its variance, for 
$D=5$ and for $p_1 = p_{c,\rm bond}$. The mass density is defined per unit area of the base surface. Both 
plots show very clearly a phase transition at $p_2 = p_{2,\rm sponge} = 0.2586(5)$.
Notice that $p_{c,\rm bond}=0.11817$ for $D=5$, i.e. there is a rather wide region with a sponge phase. 
For $p_2 >  0.2586$ the first cluster is so dense that the second one cannot penetrate and there is no 
sponge phase. 

Similar plots were made for many more values of $p_1$ and for all dimensions between 3 and 7. Final results 
for $D\leq 5$ are plotted in Fig.~1 together with the curves for the percolation (or pinning) transitions.

\begin{figure}
\begin{center}
   \includegraphics[width=0.49\textwidth]{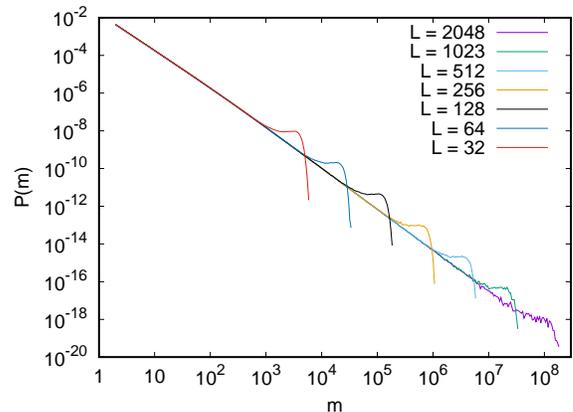}
\end{center}
        \caption{(color online) Mass distribution of clusters inside the holes of the giant cluster in supercritical
	bond percolation in $D=3$, with $p= p_{b,\rm sponge} =0.31958$ chosen such that the holes just contain 
	a incident giant cluster (notice that this second cluster is grown with $p=1$, i.e. we are at the 
	threshold where a giant cluster exists in the complement of the first cluster. As in ordinary 3-D 
	percolation, we have $P(m) \sim m^{-\tau}$ with $\tau=2.189$ in the scaling region and a peak
	position scaling $m_{\rm max} \sim L^{D_f}$ with $D_f = 2.523$.}
	\label{fig:hull_3D}
\end{figure}

Notice that it would not be easy to obtain the order of the sponge transition or any critical exponents
from plots like Fig.~2. Partly this is because these plots depend strongly on $L_z$. For $L_z \to\infty$,
not only the variance but also the average mass would diverge. But fortunately, the transition can be 
studied much more precisely by another type of simulation. 

There we use a lattice of size $L^D$ with 
periodic boundary conditions \cite{footnote2}. In a first step we determine the percolating cluster, if 
there is one. Since we do this deep in the supercritical region, this is practically always the case, 
and its determination is easy. In a second step we look for a percolating cluster with $p_1=p_2=1$ on its 
complement. Notice that we cannot show in this way 
whether the second cluster penetrates densely the first one, but we can determine the transition point 
with higher precision and we can study the second cluster more carefully near the transition. A typical result 
is shown in Fig.~3, where we plotted the mass distributions of all clusters in the holes of the first 
(supercritical) cluster for bond percolation at $p = p_{b,\rm sponge}$ in $D=3$. They have the standard 
Fisher exponent $\tau = 2.189$ 
and the fractal dimension $D_f = 2.523$ of ordinary percolation. Similar plots for dimensions $D=4$ to 7 show
that the problem is always in the ordinary percolation universality class. This might not be obvious in
view of the fact that percolation inside holes of critical percolation clusters in $D=2$ is in a different
universality class \cite{Hu}. But it actually is not very surprising, since the first 
cluster -- being strongly supercritical -- is very similar to the set of all black sites, and thus 
the second largest (white) cluster grows on a very weakly correlated set of randomly chosen sites. This
is very different in $D=2$.
A list of sponge transitions for bond percolation is given in Table~1. These values should be compared
to the asymptotic estimate 
\be
    p_{b,\rm sponge} \sim \log(D) / 2D\quad
\ee
for large $D$ \cite{Damron0}.

\begin{table}
\begin{center}
	\caption{Sponge transition points for bond percolation in dimensions $D = 3$ to 7. In the last
	column we give the thresholds for directed percolation on hypercubic lattices with the diagonal
	as preferred direction \cite{Wang2013}.}
\label{table-spongetransition}
\begin{tabular}{|l|c|c|} \hline \hline
	  D  &  $p_{b,\rm sponge}$ & $p_{b,\rm dirperc}$\\ \hline
	  3  &    0.31958(1)       & 0.38222\\
	  4  &    0.27289(1)       & 0.26836\\
	  5  &    0.24081(1)       & 0.20792\\
	  6  &    0.21582(1)       & 0.17062\\
	  7  &    0.19671(2)       & 0.14509 \\ \hline\hline
\end{tabular}
\end{center}
\end{table}

As seen from Fig.~1, there is a striking qualitative difference between $D=3$ and $D\geq 4$: While 
there is a spongy phase for all values of $p_1$ when $D\geq 4$, no such phase exists for  $D=3$ when 
$p_1 < p_{1, \rm tricrit}$. Indeed, within the limits of accuracy, the spongy phase seems to exist 
precisely down to $p_1 = p_{1, \rm tricrit}$. We have no theoretical explanation for this. It means 
that in $D=3$ the depinning transition always leads to a non-spongy growing cluster when the pinned 
interface is non-fractal, but always leads to a spongy growing cluster when it is percolation-like. 
There is no such distinction for $D\geq 4$.

\section{Are growing interfaces in media with frozen pinning centers in the KPZ universality class?}

\subsection{General remarks and a second morphological transition related to directed percolation}

When there is a sponge phase, the answer to the above question is, at least in naive sense, ``no". The 
true interface is in this case extremely convoluted, has fractal dimension $D_f = D$, and penetrates
densely the entire `bulk' phase underneath 
the surface. One can presumably define then an `outer interface' where all fjords are cut off, in which 
case one would follow what would {\it look} like the interface in a coarse grained sense.
Then we still have several obvious questions: 
\begin{itemize}
	\item If we are not in a spongy phase, is the interface scaling described by a single 
		universality class or are there different universality classes?
	\item Is at least one of these the KPZ universality class?
	\item If there are several classes, are there further morphological transitions between them?
	\item Does there exist a ``natural" definition of an {\it effective} interface in the spongy 
		phase where fjords are (at least partially) cut off, and does KPZ scaling hold for it? 
\end{itemize}

A partial answer is suggested by the fact that the KPZ equation has weak and strong coupling solutions, 
but only for $D > 3$ \cite{Nattermann,Krug-Spohn}. The strong coupling solution is the one that shows 
the standard KPZ scaling, while interfaces in the weak coupling regime are asymptotically flat with 
logarithmic corrections, i.e. the interface 
width increases less fast than any power of the base length $L$, in the limit $t\to \infty$. A similar 
weak to strong coupling transition might occur in our model already for $D=3$. Indeed, when $p_1$ is 
very small, the growth at tips of the interface is very strongly suppressed (maybe stronger than by 
any finite diffusion constant in the KPZ equation), and the interface can grow only, if bays are filled 
up sufficiently fast -- leading thereby to a non-rough interface asymptotically.

Another partial answer is also suggested by the fact that yet another morphological transition for 
interface growth with bounded speed of spreading is provided by the threshold of directed percolation
(DP) \cite{Kertesz, Krug,Krug-Spohn}. Consider, more precisely, bond percolation from an initial wetted 
surface in $D\geq 3$ with Miller indices $(1,\ldots,1)$, such that growth is along the space diagonal. 
Denote by $p_{c,\rm bond}$ the bond percolation threshold in $D$, and by $p_{b,\rm dirperc}$ the 
threshold for directed bond percolation on a $D-$dimensional hypercubic lattice with growth in the 
$(1,\ldots,1)$ direction. 
As long as $ p_{c,\rm bond} < p < p_{b,\rm dirperc}$, a typical shortest path to a wetted site with 
$z\gg 1$ will have many back-turns, simply because long paths without back-turns would not exist for 
$p < p_{b,\rm dirperc}$. This is, however, no longer true for $p> p_{b,\rm dirperc}$. In this regime,
{\it many} such paths exist, and there is a non-zero probability that a site is wetted by such a path.
Thus there is a finite probability that sites with $z=t$ get wetted at time $t$, and the active sites
are on a perfectly flat surface -- although the latter has fjords and holes penetrating it nearly 
everywhere. If we would cut off these fjords by some ad hoc rule, we would obtain a flat interface.

This last argument rests on the assumption that we have a regular lattice with frozen percolative
disorder. It would no longer hold, if we had in addition annealed disorder (e.g. with random times
needed to jump over a lattice bond), or if the frozen disorder were of a different type. In that 
case, there would still be a directed percolation transition at which back-bending paths become 
irrelevant, but paths without back-bending would not arrive at the same hight at any given time.
Thus the interface would still be rough, and the standard arguments for KPZ scaling in cases with 
annealed disorder (in particular the relationship with directed polymers \cite{Krug-Spohn,Halpin})
would suggest that such interfaces {\it are} in the KPZ universality class.

Thus we should not expect the DP-related morphology transition to be universal, while we do expect
the sponge transition to be universal. Nevertheless, it is of interest to look at the numerics.
Critical vales for DP on hypercubic lattices with spreading direction along the diagonal 
\cite{Wang2013} are given in Table 1. We see that $p_{b,\rm dirperc} > p_{b,\rm sponge}$ for $D=3$, 
but the opposite is true for $D>3$. Thus, we have no regime where we can expect true KPZ scaling of 
supercritical bond percolation interfaces in dimensions $D\geq 4$, there is a possible window 
($p\in [0.3196,0.3822]$ for bond percolation) for $D=3$. Notice, however, that we can still have
KPZ scaling for {\it effective} interfaces in higher dimensions, a question which we will not study
in the present paper.

In the following we shall study only the case $D=3$, in a wide parameter range. These will include 
spongy and non-spongy phases, and weak and strong coupling. Although our results are affected by very 
large finite size corrections, we shall argue that a weak coupling phase exists and that KPZ scaling 
holds at least approximately in the strong coupling regime for effective interfaces -- even in the 
spongy phase.

\subsection{Supercritical interfaces in $D=3$}

In this subsection we show simulations for the minimal model in $D=3$.
We use plane seeds oriented in the $(1,1,1)$ direction, i.e. with the normal to the plane parallel
to the space diagonal of the lattice. This model bears some resemblance to the cube-stacking model
of interface growth \cite{Forrest}, but with three important differences:
\begin{itemize}
	\item The cube-stacking model is not a percolation model but of Eden \cite{Eden} type, i.e.
		there are no dead (or `removed') sites, and growth at any site can occur at any time, 
		although it had stopped temporarily before.
        \item In the cube-stacking model the interface can also recede locally, by cubes detaching 
		from the interface.
	\item In the cube-stacking model, sites can be wetted only if they have three wetted neighbors,
		while we allow growth already for one or two wetted neighbors. This can create holes 
		and overhangs, which are not possible in the cube-stacking model. 
\end{itemize}

With such an orientation, interfaces can grow even if $p_1 = 0$,
i.e. even if two neighbors are needed to wet a site. Lateral b.c. are helical, but for most efficient 
use of the memory we used tilted coordinates $i$ for the horizontal and $z$ for the vertical position, 
such that the neighbors of site $(i,z)$ are $(i,z+1), (i,z-1), (i+1,z+1),(i-1,z-1),(i+L,z+1),(i-L,z-1)$. 
If we had used the original tilted coordinates, every second memory location would be unused. 

In the $z$ direction, lattices were de facto infinitely large, because 
we used a recycling trick: First, we determined in test runs the roughness in worst cases. Assume that
we estimated that the width -- for the given lateral size L and for the given values of $p_1$ and $p_2$ 
-- is less than $W_{\rm max}$ with high probability. We introduced a height variable $z_0$ and initiated 
it as $z_0=0$. Then, if the highest site in the interface had 
reached a height $z_{\rm max} > W_{\rm max}+z_0$, we erased the plane $z=z_0$, increased $z_0$ by 1
unit, and used the erased plane to store the parts of the interface with $z=z_{\rm max}$.
During these simulations we checked
that $W$ indeed remained bounded by $W_{\rm max}$. If this was violated, we discarded the entire 
simulation and repeated it with a higher value of $W_{\rm max}$. 

As a last trick to use memory most efficiently, we used multispin coding. The system sizes that we could 
handle in this way depend on the interface roughness and thus also on the distance from the critical 
percolation line. For small roughnesses, i.e. far above the percolation threshold, we could simulate 
systems with base surfaces up to $L\times L= 8192\times 8192$ and followed them typically for $2\times 
10^5$ time steps (for smaller base surfaces, up to $L=2048$, we went up to $10^6$
to $10^7$ time steps). This is to be compared to the largest previous simulation 
of interfaces with overhangs (in the RFIM) , where $t \approx L\leq 250$ \cite{Roters}.

\subsubsection{Interface width scaling for $t\to \infty$}

For small times, KPZ scaling is masked by the local roughness of the interfaces, and studying it 
numerically requires particular care. We postpone this to the next sub-subsection, and present here 
only data for large times, when the interface moves with constant
average speed and its statistical properties are stationary. Since our interfaces had overhangs and 
the bulk phase below had holes, the definition of an interface height is not unambiguous. For 
simplicity we show results where the average height and its variance at time $t$ are given by the 
average height and variance of the active (i.e., newly wetted) sites. We verified however that similar 
results were also obtained by other definitions, e.g. if the (effective) interface height at horizontal
position $i$ is given by the highest wetted site in the column $\{(i,z); z\geq 0\}$.

\begin{figure}
\begin{center}
   \includegraphics[width=0.49\textwidth]{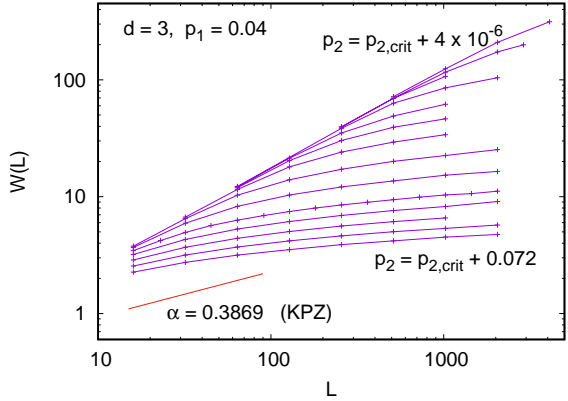}
\vglue -2.0cm
\end{center}
        \caption{(color online) Interface widths in the stationary state for large times versus base size $L$. The 
	short red straight line represents the KPZ scaling $W \sim L^{\alpha}$ with $\alpha = 0.3869$. Each of the 
	other curves is for one value of $p_2$, with $p_1=0.04$ being common to all of them. The lowest curve
	is the one for largest $p_2$, while the uppermost curve is for $p_2$ just slightly above the percolation
	threshold, i.e. for interfaces that are nearly pinned. More precisely, the values of $p_2$ are (from
	top to bottom) $0.52432, 0.52436, 0.5244, 0.52452, 0.5248, 0.5252, 0.526, 0.528,$
	$ 0.532, 0.54, 0.548, 0.56, 0.576$, and $0.596$.}
	\label{fig:widths_p_04}
\end{figure}

For KPZ in 3 dimensions, the width of a moving interface in the stationary state scales as 
\be
    W(L) \sim L^\alpha        \label{Eq.-alpha}
\ee
with $\alpha = 0.3869\pm 4$ \cite{Pagnani}. In Fig.~4 we compare this (short straight line) to data for 
$p_1 = 0.04$. Each curve here corresponds to one value of $p_2$. The uppermost curves essentially show the 
scaling of nearly pinned interfaces. The upper envelope gives $W\sim L^{0.84}$ for just marginally unpinned 
interfaces. All curves are convex. If this convexity prevails also for larger $L$, each curve gives an upper 
bound for $\alpha$. Indeed, the slopes of all curves seem to become the same, giving the bound
$\alpha \leq 0.10(1)$, in striking disagreement with the KPZ prediction.

\begin{figure}
\begin{center}
   \includegraphics[width=0.49\textwidth]{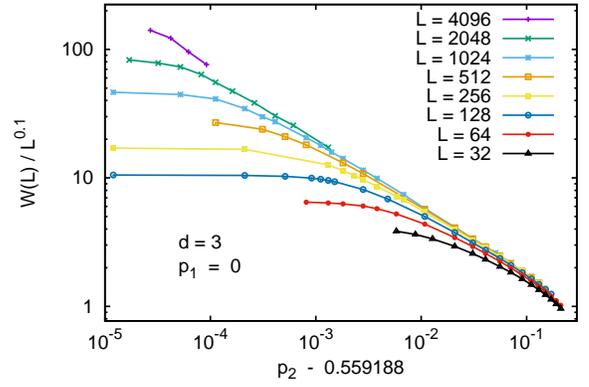}
\vglue -2.0cm
\end{center}
        \caption{(color online) Interface widths in the stationary state for large times at $p_1 = 0$ versus
	$p_2-p_{2,crit}$. Each curve corresponds to one fixed value of $L$. Anticipating that 
	$W(L) \sim L^\alpha$ with $\alpha \leq 0.1$ for large $L$, we actually plotted $W(L)/ L^{0.1}$.}
        \label{fig:widths_p_0}
\end{figure}

In order to see whether this depends on the particular value of $p_1$, we show in Fig.~5 analogous results
for $p_1 = 0$. We plot there the data differently, to indicate also the scaling for small $p_2 -p_{2,crit}$,
where $p_{2,crit} = 0.559188$ is the pinning threshold. More precisely, we plotted  in Fig.~5 the ratios
$W(L)/ L^{0.1}$,
anticipating that $W(L)$ scales also for $p_1=0$ with a similarly small exponent $\alpha$ as for $p_1=0.04$. 
This seems to be indeed the case, and again our results are in striking disagreement with the KPZ prediction. 

\begin{figure}
\begin{center}
   \includegraphics[width=0.49\textwidth]{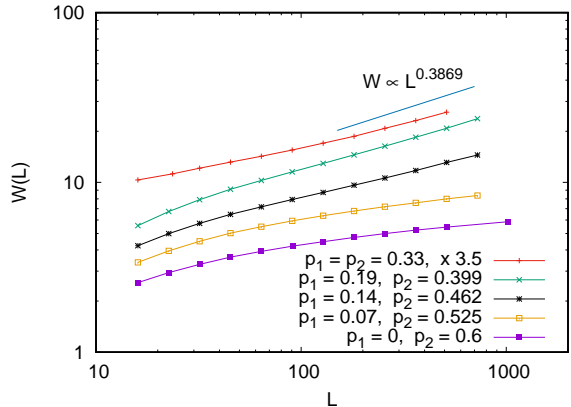}
\vglue -2.0cm
\end{center}
	\caption{(color online) Interface widths in the stationary state for large times at five pairs 
	($p_1,p_2)$ versus base size. The curve for bond percolation ($p_1 = p_2$) is shifted up to avoid 
	crowding of the curves. The short straight line indicates again KPZ scaling.}
        \label{fig:KPZ}
\end{figure}

These results might suggest that there is indeed another non-trivial power law, in addition to the KPZ 
scaling and to the trivial scaling $W=const$ of the weak-noise solution of the KPZ equation. We consider
this, however, as extremely unlikely and interpret the data in Figs.~\ref{fig:widths_p_04} and 
\ref{fig:widths_p_0} as logarithmic growth of $W(L)$, as is indeed expected for a weak coupling phase.

\begin{figure}
\begin{center}
   \includegraphics[width=0.49\textwidth]{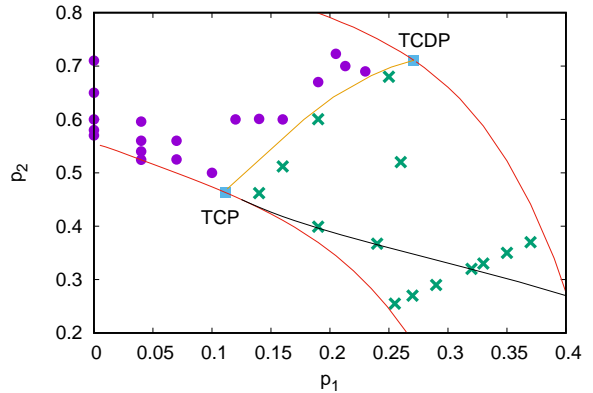}
\vglue -2.0cm
\end{center}
        \caption{(color online) Phase diagram for $D=3$ with weak coupling KPZ points indicated as bullets 
	and strong coupling points indicated as crosses. The weak/strong classification is done on the 
	basis of the curvatures of $W(L)$ versus $L$ curves (such as those in Fig.~\ref{fig:KPZ}) for 
	$L> 128$. The uppermost (red) curve indicates critical DP, with tricritical DP marked (like 
	tricritical percolation as well) by a square (light blue). The curve running SW to NE (dark yellow 
	in online version) is a rough estimate of the weak/strong transition curve. It ends on both sides 
	at tricritical points.}
        \label{fig:w-s}
\end{figure}

Results for larger values of $p_1$ are shown in Fig.\ref{fig:KPZ}. They are all for values of $p_2$ which 
are slightly ($\approx 5 \%$) higher than the critical ones, except for those for bond percolation. There,
$p$ had to be substantially larger than $p_c$ because we wanted it to be outside the sponge phase. To avoid 
crowding of curves, we thus shifted the curve for bond percolation by multiplying $W$ by a factor $3.5$.
We see that curves for $p_1 < 0.1$ bend down for all $L$, while curves for $p_1>0.1$ veer up for large $L$.
Although none of the curves reaches the KPZ scaling at large $L$ (for this we would need much larger 
system sizes), we interpret this as an indication that the transition between weak and strong coupling
happens near $p_1 \approx 0.1$. This is also supported by Fig.\ref{fig:w-s}, where we show again the $p_1$ 
versus $p_2$ plane, and indicate by bullets (crosses) points in the weak (strong) coupling regime. All 
this suggests a weak/strong transition near the short line in Fig.\ref{fig:w-s}. It is very tempting 
to suggest that this line meets the critical line precisely at the tricritical point. This fascinating 
conjecture would mean that four different transition lines meet at this point: The critical curve for 
self-affine pinned surfaces, the critical percolation curve, the sponge transition, and the weak/strong 
KPZ transition. It would also explain why the very careful RG study of Janssen {\it et al.} \cite{Janssen}
disagreed so dramatically with the numerics of \cite{Bizhani}. At its upper end, the weak/strong transition
curve ends on the critical DP curve, more precisely it seems to end at the tricritical DP point. The latter
is identified by the tricritical scaling for the density of wetted sites, $\rho(t) \sim t^{-0.087\pm 0.003}$
and the survival probability of a point seed cluster, $P(t) \sim t^{-1.218\pm 0.007}$
\cite{Grass-TCDP} (which gives $(p_1,p_2)_{\rm TCDP} = (0.2709(3),0.7111(5))$).

In the strong coupling regime there are (as in the regime with weak coupling) huge corrections to scaling, 
which makes a precise determination of $\alpha$ very difficult. The best estimates, resulting from the
points $(p_1,p_2) = (0.26,0.52), (0.24,0.3672), (0.32,0.32)$ and (0.33,0.33), are $\alpha = 0.37(2)$.
This suggests that the strong coupling regime is in the KPZ universality class, although there is also
room for caveats. 

\subsubsection{Time dependence of interface widths}

\begin{figure}
\begin{center}
	\includegraphics[width=0.49\textwidth]{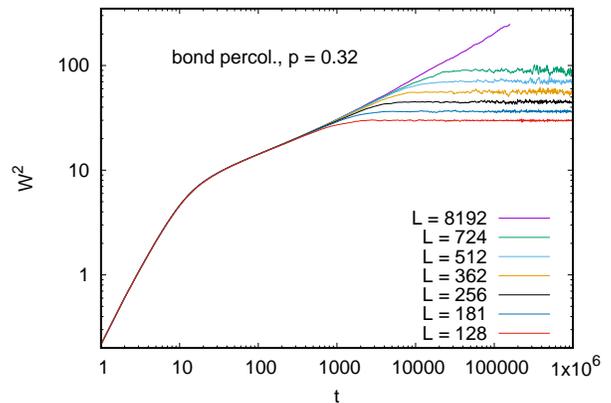}
\vglue -2.0cm
\end{center}
        \caption{(color online) Log-log plot of squares of effective interface widths for bond percolation 
	at $p = 0.32$. Each curve corresponds to one value $L$ of the system size.} 
        \label{fig8}
\end{figure}

Typical results for bond percolation at $p = 0.32$, i.e. in the strong coupling region slightly above 
the sponge transition, are shown in Fig.~8. We see essentially three regimes: 
\begin{itemize}
    \item For very short times ($t < 10$ in the present case) the widths are dominated by the fuzziness 
	of (near-critical) percolation. In this regime there is no dependence on $L$, and -- as we 
	shall see later -- also no dependence on $p$. 
    \item In the intermediate time regime ($10 < t < 10^3$ in the present case) the ``microscopic"
	(percolative) roughness is overtaken by the KPZ-type roughness, but times are still so small
	that the considered lattices are effectively infinite. Thus there is still no dependence on $L$,
	but -- as we shall see later -- there is dependence on $p$.
    \item Finally, there is the large-$t$ regime where the width no longer can grow. Rather, interfaces
        are there (statistically) stationary. This is the regime studied in the previous sub-subsection.
\end{itemize},

In Fig.~8. we included also a curve for $L=8192$, for which the stationary regime is not yet seen. We 
did this in order to make clear that the curves are not simply straight lines in the intermediate
regime. This will become even more obvious in the next two figures. 

\begin{figure}
\begin{center}
        \includegraphics[width=0.49\textwidth]{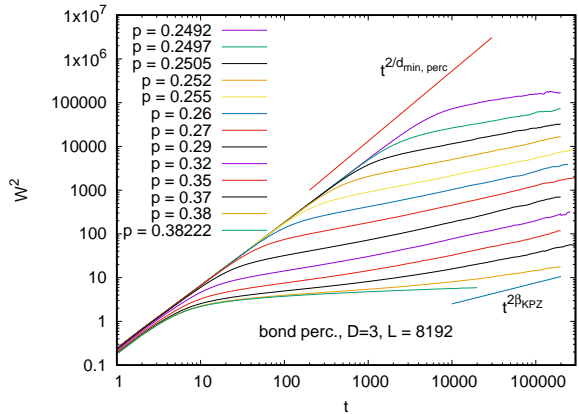}
\vglue -2.0cm
\end{center}
        \caption{(color online) Log-log plot of squares of effective interface widths for bond percolation
        at $L=8192$ and different values of $p$. Notice that all values of $p$ are supercritical, but some
	are in the sponge and others are in the non-sponge phase. The lowest curve (with the largest value
	of $p$) is for the DP-transition, while the uppermost curve is just barely supercritical. 
	Nothing special is seen near the sponge transition, which is at $p\approx 0.32$.
	The two straight lines are the scalings expected for $p = p_{b,\rm crit}$ and for KPZ.}
        \label{fig9}
\end{figure}

\begin{figure}
\begin{center}
        \includegraphics[width=0.49\textwidth]{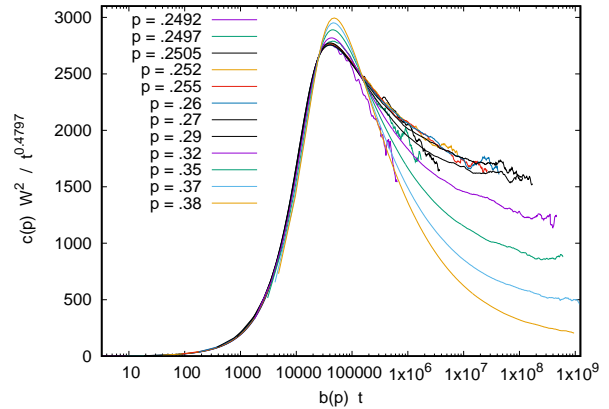}
\vglue -2.0cm
\end{center}
        \caption{(color online) Same data as in Fig.~9, but curves are shifted horizontally and vertically, 
	and divided by
        $t^{2\beta_{\rm KPZ}}$. Plotted in this way it becomes obvious that systems are too small for 
	$p < 0.251$ to show a possible KPZ scaling, while $t$ is too small for $p>0.28$.}
        \label{fig10}
\end{figure}

In Fig.~9 we show again $W^2$ versus $t$ (again for bond percolation), but now we keep $L=8192$ fixed, 
and vary $p$. Also, due to the very large $L$ all data in Fig.~9 are in the small- and intermediate-$t$ 
regimes. We compare these data with two power laws indicated by straight lines. The straight line on 
the left indicates the behavior for critical
percolation. It does not seem to fit perfectly, but this is (in spite of the large value of $L$) a 
finite-size effect. Closer scrutiny shows that the data are there in perfect agreement with theory. 
The straight line on the r.h. side corresponds to what is expected for KPZ scaling. There the agreement 
is not as good. For the times shown in this figure, all curves are more flat than predicted for KPZ, 
but this is maybe due to the curvature seen already in the previous figure. We cannot rule out that the 
curves would for larger $t$ and thus also for larger $L$) slowly approach the KPZ prediction. To make this 
even more clear, we plotted the same data in a slightly different way in Fig.~10, where we multiplied each 
curve by $t^{-2\beta_{\rm KPZ}}$ and shifted it horizontally and vertically, in order to enhance the
significance of the plot.

We should stress that Figs~9 and 10 also include data from the sponge phase, where the true interface
width would increase linearly with $t$. What these data show is that our definition of effective interfaces
cuts off fjords sufficiently, so that they have even a good chance to satisfy KPZ scaling.

Finally, we should mention that data in the weak coupling regime look qualitatively similar to Figs.~8 and 9,
with three regimes dominated by internal roughness (for small $t$), by logarithmic increase of $W$ 
(intermediate $t$) and corresponding to asymptotic stationarity. We do not show these data because they no 
not look very distinct.

\section{Discussion and conclusions}

The main result of the present paper is the establishment of a sponge phase. Similar phases with two 
tightly intermingled microscopic phases were previously known only for systems with at least three 
microscopic phases, e.g. in emulsions with two liquids and one membrane \cite{Roux}. In the present case,
there are no explicit membranes between the two microphases, which are made up of interpenetrating 
supercritical percolation clusters. The striking property which distinguishes a sponge phase from other 
multiphase systems is that both microphases are connected and everywhere dense in the mathematical sense,
i.e. in the scaling limit every point in space is infinitely close to points in both microphases.

This sponge phase implies that the standard scenario for interfaces moving in locally isotropic media with
quenched randomness, going back to a famous paper by Bruinsma and Aeppli in 1984 \cite{Bruinsma}, is wrong. 
It basically says that such interfaces become more and more smooth as dimension increases. While this is 
correct for strongly pushed interfaces far from local pinning, the opposite is true close to the pinning
transition: There, moving interfaces become more and more fuzzy with increasing dimension. What is however
still true -- at least for $D=3$ -- is that sites where the interface {\it moves} are located on a smooth
manifold, even when the interface itself is not smooth.

For $D>3$ we found that the transition from pinned to moving interfaces in media with frozen pinning centers
always leads first to a sponge phase, while non-spongy bulk phases appear only in a second morphological
transition. This is at least what we found in what was called the `minimal model' in \cite{Grass2018}, but 
we conjecture that the same is true also for the RFIM with any other distribution of local random fields.

For $D=2$ no sponge phases can exist (for topological reasons), and for $D=3$ we found that the de-pinning
transition can lead either to a sponge or to a non-sponge phase. The first seems to happen when the 
de-pinning transition is in the percolation universality class, while the latter happens when the critically
pinned interface is self-affine. Again, this claim is only based on the minimal model. Simulations with 
other models in the generalized percolation family (which includes the zero-T RFIM) would be extremely 
welcome to verify it, as would be analytic arguments. So far, these claims are entirely based on simulations.

At the sponge / non-sponge transition, the incipient infinite clusters in the voids of the main phase
are in the universality class of critical ordinary percolation. Again we claim that
this is true in general, although we have only simulation data to support it.

In the last part of the paper, we asked whether the interfaces in the parameter region above the sponge
phase satisfy KPZ scaling, as is often assumed -- in spite of the fact that theoretical arguments for 
KPZ scaling only exist in cases without frozen pinning centers. Moreover, we asked whether suitably defined 
{\it effective} interfaces could satisfy KPZ scaling even in the sponge phase. Here we studied only the case 
$D=3$. We found huge finite size corrections, but our simulations were nevertheless able to support the 
following scenario: For all control parameters, the interfaces are asymptotically either in a strong-coupling
universality class, or in the weak-coupling class. In the latter, the interface width increases, for 
large times, less fast that any power of the base length $L$, so that it becomes asymptotically flat.

Whether the strong coupling universality class is really identical with the KPZ class is still an open 
problem. Usually it is taken for granted, but without any compelling theoretical or numerical arguments.
In all present simulations, corrections to scaling were extremely large, so that no firm conclusion
could be drawn. But the values of the exponents $\alpha$ and $\beta$ found in the strong coupling domain 
suggest that moving rough interfaces in media with frozen pinning centers might indeed be in the KPZ class.

Marginally supercritical (i.e., nearly pinned) interfaces in $D=3$ seem to be rough only when 
the {\it critically pinned} interfaces are percolation hulls in the standard percolation class, while 
de-pinning via self-affine critical interfaces leads to asymptotically flat interfaces (weak coupling 
KPZ phase). This might not be the most important finding of the present paper, but it certainly is the 
most surprising -- together with the observation that the weak / strong coupling transition curve 
ends also at its other end at a tricritical point, namely that of DP. The fact that the weak / strong 
coupling transition curve ends on {\it both} sides at tricritical points suggests strongly that this is 
not merely a numerical accident.

To many readers, the very existence of a weak coupling (i.e., non-rough) phase in $D=3$ will be the most 
surprising result. It presents a clear violation of the often assumed KPZ universality. Reasons for 
possible non-universality have been discussed in the literature \cite{Newman,Tripathy,Wolf,Amar,Huse,Krug2}, 
but none of these seems to apply in the 
present case. They include unusual noise distributions in high dimensions \cite{Newman}, pulled fronts
\cite{Tripathy}, anisotropies in the KPZ non-linearities \cite{Wolf}, and vanishing of the KPZ nonlinearity 
parameter \cite{Amar,Huse,Krug2}. 

Although our noise looks unconventional in the RFIM interpretation, it is very similar to a simple Gaussian 
\cite{Grass2018}, and according to \cite{Newman} we should expect problems anyhow only for $D>3$. In the 
terminology of \cite{Tripathy}, our fronts look
much more like pushed than like pulled fronts, in particular for small $p_1$ where we see the weak coupling
phase. Our model is anisotropic in the sense that the speed of propagation of a tilted interface
depends on the orientation of the tilt. But it has a sixfold symmetry in the sense that the speed is lowest
when the tilt makes the interface close to a coordinate plane, while it is maximal when it is in the 
direction of one of the six diagonals. In contrast, an effect was found in \cite{Wolf} only for two-fold
symmetries. Finally, there is no indication in our model that the KPZ nonlinearity vanishes, in contrast to
the model of \cite{Amar}. Also, in the latter the weak coupling 'phase' is, strictly spoken, observed 
only at an isolated parameter value, while we observe a true phase in an extended parameter region.

In contrast to these effects, we believe that the origin of the weak-coupling phase (and thus of the 
non-universality with KPZ) is that, for small $p_1$, the suppression of peaks and spikes in the 
interface is stronger than what can be described by the standard diffusion term in KPZ. Phenomenologically,
this might be described by a generalization of KPZ where a lower power power of the Laplacian is added, 
e.g. in Fourier space $\partial_t h = -\nu'k^\mu h -\nu k^2 h + \frac{1}{2} \lambda ({\bf k} h)^2 + \eta$
with $\nu'>0$ and $0<\mu<2$ \cite{Nicoli1,Nicoli2}.

Since the present work is mainly numerical, and moreover based only on one particular model, the most 
obvious open problems are to support our claims with mathematical arguments, and to do simulations 
for more general model classes. Apart from that, there are several minor open problems, such as the
behavior near the tricritical point in $D=3$. Also, there are several problems we have not yet mentioned.
For instance,
at the sponge transition for bond percolation a supercritical cluster with $p>p_c$ coexists with a
critical cluster with $p=1$. A more general problem would be to find the region(s) in control parameter
space where one giant cluster with $p=p_1$ coexists with another one with $p=p_2$.

  \begin{acknowledgments}
	  I thank Michael Damron for very helpful correspondence, and to Joachim Krug, Wolfhard Janke, 
	  Werner Krauth, and Gerhard Gompper for discussions. I also want to thank Maya Paczuski and 
	  Golnoosh Bizhani for collaborations during very early step of this work, and the Complexity Science
	  Group at the University of Calgary as well as the Max-Planck Institute for the Physics of Complex 
	  Systems, Dresden, for generous grants of computer time.
  \end{acknowledgments}


\begin{thebibliography}{99}
\bibitem{footnote1} Actually, most simulations were done on tilted simple (hyper-)cubic lattices, such that the z-axis
        is parallel to one of the space diagonals. In that case, all neighbors of a point $(\ldots,z)$ have $z' =
        z\pm 1$.
\bibitem{footnote2} Actually we used, for easier coding and higher speed, helical boundary conditions on lattices
	with $2^k$ sites, such that $L\approx 2^{k/D}$. The difference with strictly periodic b.c. is negligible.
\end{thebibliography}
\end{document}